\documentclass[12pt,twoside,fleqn]{article}

\usepackage{a4wide,cite}
\usepackage{amsmath,amssymb}
\usepackage{epsfig,rotating}
\usepackage{axodraw}

\numberwithin{equation}{section}
\allowdisplaybreaks[4]

\def \be{\begin{equation}}
\def \ee{\end{equation}}
\newcommand{\bea}{\begin{eqnarray}}
\newcommand{\eea}{\end{eqnarray}}
\def \nn{\nonumber}

\newcommand{\MS}{\ensuremath{\overline{\text{MS}}}}

\begin{document}

\begin{titlepage}

\begin{flushright}
  SFB/CPP-10-39\\
  MZ-TH/10-14\\
  TTP10-26
\end{flushright}

\vskip 2cm

\centerline{\Large\bf\boldmath Complete $(O_7,O_8)$ contribution to $\bar B\to X_s\gamma$ at $O(\alpha_s^2)$}
\vskip 1cm

\begin{center}
  {\bf H.M.~Asatrian$^a$, T.~Ewerth$^{b,c}$, A.~Ferroglia$^{d,f}$, C.~Greub$^e$, and G.~Ossola$^f$}\\[1mm]
  {$^a$\sl Yerevan Physics Institute, 0036 Yerevan, Armenia}\\[1mm]
  {$^b$\sl Institut f{\"u}r Theoretische Teilchenphysik, Karlsruhe Institute of Technology (KIT),\\
    D-76128 Karlsruhe, Germany}\\[1mm]
  {$^c$\sl Dip.\ Fisica Teorica, Univ.\ di Torino \& INFN Torino, I-10125 Torino, Italy}\\[1mm]
  {$^d$\sl Institut f{\"u}r Physik (THEP), Johannes Gutenberg-Universit{\"a}t\\ D-55099 Mainz, Germany}\\[1mm]
  {$^e$\sl Albert Einstein Center for Fundamental Physics, Institute for Theoretical Physics,\\
    Univ.~of Bern, CH-3012 Bern, Switzerland}\\[1mm]
  {$^f$\sl Physics Department, New York City College of Technology, 300 Jay Street,\\
    Brooklyn NY 11201, USA}
\end{center}
\medskip

\vskip 1cm

\begin{abstract}
\noindent 
We calculate the set of $O(\alpha_s^2)$ corrections to the branching ratio
and to the photon energy spectrum of the decay process 
$\bar B\to X_s\gamma$ originating from
the interference of  diagrams  involving the electromagnetic dipole operator
$O_7$ with diagrams involving the  chromomagnetic dipole operator  $O_8$.
The corrections evaluated here are one of the elements needed to complete
the calculations of the $\bar B\to X_s\gamma$ branching ratio at
next-to-next-to-leading order in QCD.
We conclude that this set of corrections does not
change the central value of the Standard Model prediction
for $\text{Br}(\bar B\to X_s\gamma)$ by more than $1\%$.
\end{abstract}

\end{titlepage}


\section{\hspace{-3mm}Introduction}

The first estimate of the $\bar{B}\to X_s\gamma$ branching ratio within the Standard Model at the
next-to-next-to-leading order (NNLO) level was published some years ago \cite{Misiak:2006zs}:
\begin{equation}\label{estimate}
  \text{Br}(\bar B\to X_s\gamma)_{{\rm SM},\,E_\gamma > 1.6\,{\rm GeV}} = (3.15\pm 0.23)\times 10^{-4}\,.
\end{equation}
This estimate combines a number of different corrections which were calculated by several groups
\cite{Misiak:2004ew,Misiak:2010sk,Gorbahn:2004my,Gorbahn:2005sa,
  Czakon:2006ss,Blokland:2005uk,Melnikov:2005bx,Asatrian:2006ph,Asatrian:2006sm,Bieri:2003ue,Misiak:2006ab}.
The prediction given in Eq.~(\ref{estimate}) must be compared with the current world averages,
\begin{equation}\label{experiment}
  \text{Br}(\bar B\to X_s\gamma)_{{\rm exp},\,E_\gamma > 1.6\,{\rm GeV}} =
  \left\{\begin{array}{l}
  \left(3.55\pm 0.24\pm 0.09\right)\times 10^{-4}\,,\mbox{\;\,(HFAG) \cite{Barberio:2008fa}}\\[1mm]
  \left(3.50\pm 0.14\pm 0.10\right)\times 10^{-4}\,,\mbox{\cite{Artuso:2009jw}}
  \end{array}\right.
\end{equation}
which include measurements from CLEO, BaBar and Belle \cite{Chen:2001fja,Aubert:2007my,Limosani:2009qg}.
The central values of the theoretical prediction and of the HFAG average are compatible at the 1.2$\sigma$
level, while both the theoretical and experimental uncertainties are very similar in size (about 7\%). Since the
experimental uncertainty is expected to decrease
to 5\% by the end of the B-factory era (which is already indicated by the average given in the second line of
Eq.~(\ref{experiment})), it is also desirable to reduce the theoretical uncertainty accordingly.

Unfortunately, at this level of accuracy, the theoretical uncertainty is dominated by non-perturbative
contributions. As long as one restricts 
the analysis to processes mediated by the electromagnetic dipole
operator
$O_7=\alpha_{\rm em}/(4\pi)m_b\left(\bar s\sigma^{\mu\nu}P_Rb\right)F_{\mu\nu}$ alone, non-perturbative effects are
well under control \cite{Falk:1993dh,Bigi:1992ne,Bauer:1997fe,Gremm:1996df,Ewerth:2009yr}. However, as soon as
operators other than $O_7$ (such as the
chromomagnetic dipole operator
$O_8=g_s/(16\pi^2)m_b\left(\bar s\sigma^{\mu\nu}P_RT^ab\right)G_{\mu\nu}^a$) are involved,
one encounters
non-perturbative effects of $O(\alpha_s\Lambda_{\rm QCD}/m_b)$.
At present, the latter   can only be estimated
\cite{Lee:2006wn}. Hence a 5\% uncertainty related to all of the unknown non-perturbative effects has been
included in Eq.~(\ref{estimate}). A further reduction of the theoretical uncertainty below
the 5\% level
seems to be rather difficult \cite{Benzke:2010js}.
Still, given the importance of Br$(\bar B\to X_s\gamma)$ in constraining physics scenarios beyond the
Standard Model \cite{Domingo:2007dx}, it is worth to reduce the perturbative uncertainties
as much as possible. 

In particular, it would be desirable to reduce the 
uncertainty associated to the interpolation in $m_c$ which was employed to obtain Eq.~(\ref{estimate})
\cite{Misiak:2006ab}.
To get rid of the interpolation in $m_c$ in the calculation of the branching ratio is a highly challenging
task and it would represent a clear improvement
of the theoretical prediction.
Indeed, considering the work that has been done since the publication of \cite{Misiak:2006zs}, and the work that
is still in progress, an update of the estimate given in Eq.~(\ref{estimate}) will soon be  warranted.
Here we
would like to mention that the effects of charm and bottom quark masses on gluon lines are now completely known
(provided that  one neglects on-shell amplitudes that are proportional to the small Wilson coefficients of the
four-quark
operators $O_3$-$O_6$) \cite{Asatrian:2006rq,Boughezal:2007ny,Ewerth:2008nv,Pak:2008qt}. Therefore this part
could be removed from the interpolation. Also the $O(\alpha_s^2\beta_0)$-effects in the $(O_2,O_2)$, $(O_2,O_7)$ and
$(O_7,O_8)$-interference, which are known  \cite{Ligeti:1999ea}, were not considered in \cite{Misiak:2006zs,
  Misiak:2006ab}. Finally,
the complete calculation of the $(O_2,O_7)$-interference for $m_c=0$ is well underway \cite{schutzmeier}.
The latter calculation in particular will help to fix the boundary for the $m_c$ interpolation for vanishing $m_c$;
this in turn would allow one   to reduce
the 3\% uncertainty in Eq.~(\ref{estimate}) due to the interpolation. For complete up-to-date lists
of needed perturbative and non-perturbative corrections to the branching ratio we refer the reader to
the reviews \cite{Misiak:2008ss,Ferroglia:2008yg,Ewerth:2009yk,Misiak:2009nr}.

In this paper we calculate the complete  $(O_7,O_8)$-interference corrections at $O(\alpha_s^2)$ to the photon
energy spectrum
$d\Gamma(b\to X^{\rm partonic}_s\gamma)/dE_\gamma$ and to the total decay width
$\Gamma(b\to X^{\rm partonic}_s\gamma)|_{E_\gamma>E_0}$, where $E_0$ denotes the lower cut in the photon energy.
The contributions containing massless and massive quark loops were already
presented in \cite{Bieri:2003ue,Ligeti:1999ea} and \cite{Ewerth:2008nv}, respectively; the contributions
which are not yet available in the literature are the ones proportional to the color factors
$C_F^2$ and $C_F C_A$. From the technical point of view, the latter are the most complicated to evaluate and
are the main subject of the present work.

The paper is organized as follows: In Sec.~2 we present our results
for the (integrated) photon energy spectrum. In Sec.~3 we provide some details
about the calculation of the corrections proportional to $ \alpha_s^2 C_F^2$ 
and $ \alpha_s^2 C_F C_A$ by analyzing the contribution of a particular Feynman diagram. The numerical impact of
the
$(O_7,O_8)$ interference on the theoretical prediction for $\text{Br}(\bar B\to X_s\gamma)$ at NNLO is
estimated in Sec.~4. Finally, we present our conclusions in Sec.~5.


\section{\hspace{-3mm}\boldmath Results for the (integrated) photon energy spectrum}

Within the low-energy effective theory, the partonic $b\to X_s\gamma$ decay rate can be written as
\begin{equation}\label{decay_rate}
  \Gamma(b\to X_s^{\rm parton}\gamma)_{E_\gamma>E_0} = \frac{G_F^2\alpha_{\rm em}
    \overline{m}_b^2(\mu)m_b^3}{32\pi^4}\,|V_{tb}^{}V_{ts}^*|^2\,\sum_{i\le j}C_i^{\rm eff}(\mu)\,
  C_j^{\rm eff}(\mu)\int_{z_0}^1\!dz\,\frac{dG_{ij}(z,\mu)}{dz}\,,
\end{equation}
where $m_b$ and $\overline{m}_b(\mu)$ denote the pole and the running $\MS$ mass of the $b$ quark, respectively,
$C_i^{\rm eff}(\mu)$ indicates the effective Wilson coefficients at the low-energy scale, $z=2E_\gamma/m_b$ is
the rescaled photon
energy, and $z_0=2E_0/m_b$ is the rescaled energy cut in the photon energy spectrum.\footnote{In this paper we assume
  that the products $C_i^{\rm eff}(\mu)\,C_j^{\rm eff}(\mu)$ are real quantities. Therefore our formulas are not
  applicable to physics scenarios beyond the Standard Model which produce complex short distance couplings.}

As already anticipated in the introduction, we will focus on the function $dG_{78}(z,\mu)/dz$ corresponding to the
interference of the electro- and the chromomagnetic dipole operators
\begin{align}
  O_7 &= \frac{e}{16\pi^2}\,\overline{m}_b(\mu)\left(\bar s\sigma^{\mu\nu}P_Rb\right)F_{\mu\nu}\,,\\[2mm]
  O_8 &= \frac{g}{16\pi^2}\,\overline{m}_b(\mu)\left(\bar s\sigma^{\mu\nu}P_RT^ab\right)G_{\mu\nu}^a\,.
\end{align}
In NNLO approximation $G_{7 8}$ can be rewritten as follows,
\begin{equation}
\label{mastereq}
  \frac{dG_{78}(z,\mu)}{dz} = \frac{\alpha_s(\mu)}{4\pi}\,C_F\widetilde Y^{(1)}(z,\mu) +
  \left(\frac{\alpha_s(\mu)}{4\pi}\right)^2C_F\widetilde Y^{(2)}(z,\mu) + O(\alpha_s^3)\,,
\end{equation}
where
$\alpha_s(\mu)$ indicates the running coupling constant in the $\MS$ scheme
and 
\begin{align}
 \widetilde Y^{(1)}(z,\mu) &= \left[\frac{2}{9}\left(33-2\pi^2\right) +
   \frac{16}{3}\,\ln\left(\frac{\mu}{m_b}\right)\right]\,\delta(1-z)\nonumber\\[2mm]
 &\qquad
 +\frac{2}{3}\left(z^2+4\right)-\frac{8}{3}\left(1-\frac{1}{z}\right)\ln(1-z)\,.
\end{align}
The function $\widetilde Y^{(2)}(z,\mu)$ can  be split further 
into a sum of contributions proportional to different
color factors:
\begin{align}\label{colorf}
  \widetilde Y^{(2)}(z,\mu) &= C_F \widetilde Y^{(2,\mbox{{\tiny CF}})}(z,\mu) +
  C_A \widetilde Y^{(2,\mbox{{\tiny CA}})}(z,\mu)\nonumber\\[2mm]
  &\qquad+ T_R N_L \widetilde Y^{(2,\mbox{{\tiny NL}})}(z,\mu) + T_R N_H \widetilde Y^{(2,\mbox{{\tiny NH}})}(z,\mu) +
    T_R N_V \widetilde Y^{(2,\mbox{{\tiny NV}})}(z,\mu)\,.
\end{align}
Here, $N_L$, $N_H$ and $N_V$ denote the number of light ($m_q=0$), heavy ($m_q=m_b$), and purely virtual ($m_q=m_c$)
quark flavors, respectively; $C_F$, $C_A$
and $T_R$ are the $\mbox{SU}(3)$ color factors with numerical values
given by 4/3, 3 and 1/2, respectively. 
The expressions for the
functions $\widetilde Y^{(2,i)}(z,\mu)$ with $i=\mbox{\small NL}$, $\mbox{\small NH}$, $\mbox{\small NV}$ can be
found in \cite{Ewerth:2008nv}.
The main result of the present work  are the  so far unknown functions $\widetilde Y^{(2,i)}(z,\mu)$ with
$i=\mbox{\small CF}$,
$\mbox{\small CA}$, which  are given by
\begin{align}\label{Yabelian}
  \widetilde Y^{(2,\mbox{{\tiny CF}})}(z,\mu) &= \left(-37.1831-\frac{64}{3}\,L_\mu -
    \frac{128}{3}\,L_\mu^2\right)\,\delta(1-z) - 11.7874\,\left[\frac{\ln(1-z)}{1-z}\right]_+\nonumber\\[3mm]
  &\qquad - 20.6279\,\left[\frac{1}{1-z}\right]_+ - 41.7874\,\ln(1-z) - 6.6667\,\ln^2(1-z)\nonumber\\[3mm]
  &\qquad + f_1(z) - 12\,\widetilde Y^{(1)}(z,m_b)L_\mu + \frac{64}{3} H^{(1)}(z,m_b)L_\mu\,,
\end{align}
\begin{align}\label{Ynonab}
  \widetilde Y^{(2,\mbox{{\tiny CA}})}(z,\mu) &= \left(4.7666+\frac{808}{27}\,L_\mu +
    \frac{272}{9}\,L_\mu^2\right)\,\delta(1-z)\nonumber\\[2mm]
  &\qquad - 6.5024\,\ln(1-z) + f_2(z) + \frac{34}{3}\,\widetilde Y^{(1)}(z,m_b)L_\mu\,,
\end{align}
where
\begin{align}
 H^{(1)}(z,m_b) &= { -\left(\frac{5}{4}+\frac{\pi^2}{3}\right)}\,
 \delta(1-z) - \left[\frac{\ln(1-z)}{1-z} \right]_+\nonumber\\[1mm]
 &\qquad -{\frac{7}{4}}\left[\frac{1}{1-z} \right]_+\!\!-
 \frac{z+1}{2}\,\ln(1-z) + \frac{7+ z - 2 z^2}{4}\,, 
\end{align}
\begin{align}
  f_1(z) &= 20.6279 - 108.484\,z + 13.264\,z^2 + 16.1268\,z^3 - 33.2188\,z^4\nonumber\\[1mm]
  &\qquad + 69.8819\,z^5 - 111.088\,z^6 + 118.405\,z^7 - 79.6963\,z^8 + 29.929\,z^9\nonumber\\[1mm]
  &\qquad - 4.76579\,z^{10} - 56.8265\,(1-z)\ln(1-z)\nonumber\\[1mm]
  &\qquad - 8.11265\,(1-z)\ln^2(1-z) - 5.77146\,(1-z)\ln^3(1-z)\,,
\end{align}
\begin{align}
  f_2(z) &= 17.0559\,z + 20.9072\,z^2 - 0.471626\,z^3 + 10.1494\,z^4\nonumber\\[1mm]
  &\qquad - 17.4241\,z^5 + 24.7733\,z^6 - 20.4582\,z^7 + 8.47394\,z^8 - 0.173599\,z^9\nonumber\\[1mm]
  &\qquad - 0.657813\,z^{10} + 5.66536\,(1-z)\ln(1-z)\nonumber\\[1mm]
  &\qquad - 11.1319\,(1-z)\ln^2(1-z) + 1.3999\,(1-z)\ln^3(1-z)\,.
\end{align}
Note that function $H^{(1)}(z,\mu)$ also appeared in Eq.~(2.11) of Ref.~\cite{Asatrian:2006sm},
and that we introduced the short-hand notation $L_\mu=\ln(\mu/m_b)$.

In Eqs.~(\ref{Yabelian}) and (\ref{Ynonab}) the $z$-dependence of the $\mu$-dependent terms and of those
terms which become singular in the limit $z \to 1$ is exact. The functions $f_1(z)$ and $f_2(z)$ were
instead obtained by making an ansatz 
for our numerical results of the non-singular parts, using the 
functional form
\begin{equation}
\label{fitform}
f_i(z)=\sum_{j=0}^{10} c_{i,j} \, z^j + 
c_{i,11} (1-z) \ln(1-z) +
c_{i,12} (1-z) \ln^2(1-z) +
c_{i,13} (1-z) \ln^3(1-z) \, .
\end{equation}
The coefficients $c_{0},\dots,c_{13}$ were then determined by performing a
least-square fit, using 100 specific 'data'-points. We checked that
the fit-functions remain essentially the same when changing the set of
data-points. In particular, the integrals of the fit-functions, taken
in an interval $[z_0,1]$ $(0 \le z_0 < 1)$, remain basically
unchanged. The same holds true when changing the functional ansatz given in Eq.~(\ref{fitform}), 
e.g., to contain additional terms proportional to $(1-z)^2 \, \ln^{n}(1-z)$, with $n=1,2,3$.

The plus distributions appearing in Eq.~(\ref{Yabelian}) are defined as
\begin{equation}\label{plusdis1}
  \int_0^1\!dz\left[\frac{\ln^n(1-z)}{1-z}\right]_+g(z) =
  \int_0^1\!dz\,\frac{\ln^n(1-z)}{1-z}\left[g(z)-g(1)\right]\,,
\end{equation}
where $g(z)$ is an arbitrary test function which is regular at $z=1$, and $n=0,1$. In case the
integration does not include the endpoint $z=1$, we have ($c<1$)
\begin{equation}\label{plusdis2}
  \int_0^c\!dz\left[\frac{\ln^n(1-z)}{1-z}\right]_+g(z) =
  \int_0^c\!dz\,\frac{\ln^n(1-z)}{1-z} g(z)\,.
\end{equation}

We observe that the plus distributions are present only in the part of the spectrum proportional to $C_F^2$,
see Eq.~(\ref{Yabelian}).
This is in agreement with the results reported in \cite{Neubert:2004dd};
following the procedure presented in that work, it is possible to conclude 
that the plus distributions appearing in the $(O_7,O_8)$ component of the 
photon energy spectrum at ${\mathcal O (\alpha_s^2)}$ must be the same ones as in the $(O_7,O_7)$
component of the spectrum at  ${\mathcal O (\alpha_s)}$ (up to an overall factor). In particular
\bea
\left. \widetilde Y^{(2,\mbox{{\tiny CF}})}(z,m_b)\right|_{\mbox{{\footnotesize
plus distrib.}}}  &=& - \frac{8}{9} \left( 33 - 2 \pi^2 \right) \left\{   
\left[\frac{\ln(1-z)}{1-z} \right]_+ + 
\frac{7}{4} \left[\frac{1}{1-z} \right]_+ \right\} \, , \nn \\[2mm]
&=& - 11.7874 \left[\frac{\ln(1-z)}{1-z} \right]_+ - 20.6279  \left[\frac{1}{1-z} \right]_+ \, .
\label{plusdist}
\eea
The structure in Eq.~(\ref{plusdist}) emerges in our diagrammatic
calculation from delicate cancellations 
among several contributions, and it provides a valuable 
test for our result.


\section{\hspace{-3mm}Technical details about the calculation}

In order to obtain 
the results in Eqs.~(\ref{Yabelian}) and (\ref{Ynonab}) one needs to evaluate Feynman diagrams
contributing to the process $b \to s \gamma$ up to two loops,
one-loop Feynman diagrams contributing to the process $b \to s \gamma g$,
and tree-level Feynman diagrams  contributing to the processes $b \to s \gamma g g$ (plus corresponding
diagrams involving unphysical ghosts in the final state) and $b \to s \gamma s \bar{s}$.
As discussed in detail in \cite{Asatrian:2006sm}, the interferences among the various partonic diagrams with
2, 3 and 4
particles in the final state are in one-to-one correspondence with the 2-, 3- and 4-particle cuts of the 
three-loop $b$-quarks self-energy diagrams  shown in Figs.~\ref{fermionloop}, \ref{nonabelian}, \ref{double},
\ref{abelianCA},
and \ref{abelianCF2p1}, provided that
the cut goes through the photon propagator\footnote{We only display the diagrams that contribute to
  the functions $\widetilde Y^{(2,i)}(z,\mu)$ with $i=\mbox{\small CF},\,\mbox{\small CA}$.}.
The 2-particle  cuts of the self-energy graphs correspond to the interference of a tree-level and a two-loop
diagrams for the process $b \to s \gamma$, or to the interference of two one-loop diagrams for the same process.
The 3-particle cuts split a self-energy graph into a one-loop and a tree-level
diagram contributing   to the process $b \to s \gamma g$. Finally, 
the 4-particle cuts correspond to the interference of two tree-level diagrams for the process $b \to s g g$ or
for the process $b \to s \gamma s \bar{s}$. 
 
The contribution of each single cut to the photon energy spectrum can
be obtained by employing the Cutkosky rules \cite{cut1,cut2,cut3}. In order to keep the energy of the
photon fixed, it
is necessary to insert in the integrands a factor
\cite{Melnikov:2005bx} 
\be
\delta \left(E_\gamma - \frac{p_b \cdot p_\gamma}{m_b} \right) = 
2 m_b\, \delta \left((p_b - p_\gamma)^2 - (1-z) m_b^2\right) \,,
\label{energyfix}
\ee
where $p_b$ and $p_\gamma$ denote the four momenta of the $b$ quark and the photon, respectively. 
Finally, the delta function in Eq.~(\ref{energyfix}) and all of the delta functions originating from the
Cutkosky rules for the propagators crossed by a cut can be rewritten as differences of propagators as
follows \cite{Anastasiou:2002yz,Anastasiou:2003ds},
\be
\delta\left( q^2  - m^2 \right) = \frac{1}{2 \pi i} \left( 
\frac{1}{ q^2  - m^2 -i 0} - \frac{1}{q^2  - m^2 +i 0} \right) \, .
\ee
It is then possible to evaluate all of the relevant cuts of the three-loop self-energy diagrams by first
identifying a set of Master Integrals for each cut, 
and then by 
evaluating those Master Integrals by means of the tools and techniques usually employed in the
calculation of multi-loop Feynman integrals.
As usual, we work in $d=4-2\epsilon$ space-time dimensions to regularize ultraviolet, infrared and collinear
singularities.
In the rest of this section we discuss some features of this procedure
by considering one of the simplest Feynman diagrams we encountered in the course of the
calculation. More technical details concerning the 
parameterization of the phase-space integrals can be found in
\cite{Anastasiou:2003gr,Asatrian:2006ph,Asatrian:2006sm}. 

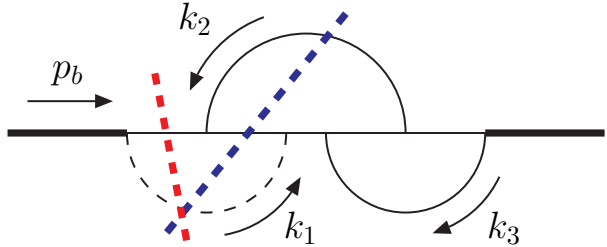
\begin{figure}[t]
\begin{center}
\vspace*{18mm}
\begin{picture}(0,0)(0,0)
\SetScale{1.5}
%
  \SetWidth{1.8}
\Line(-75,0)(-45,0)
\Line(45,0)(75,0)
  \SetWidth{.5}
\Line(-45,0)(45,0)
\DashCArc(-25,0)(20,-180,0){3}
\LongArrowArc(-25,0)(26,-80,-25)
%
%
\CArc(25,0)(20,-180,0)
\LongArrowArcn(25,0)(26,-25,-75)
\CArc(0,0)(25,0,180)
\LongArrowArc(0,0)(31,110,160)
%
%
%
\LongArrow(-70,8)(-50,8)
\Text(-90,18)[cb]{{\large$p_b$}}
\Text(-2,-42)[cb]{{\large $k_1$}}
\Text(-42,+37)[cb]{{\large $k_2$}}
\Text(75,-42)[cb]{{\large $k_3$}}
  \SetWidth{1.8}
  \SetColor{Blue}
\DashLine(-35,-25)(10,30){3}
  \SetColor{Red}
\DashLine(-30,-27)(-38,15){3}
\end{picture}
\end{center} 
\vspace*{15mm}
\caption{{\sl A three-loop self-energy topology and the corresponding 2- and 3-particle cuts. Thin lines represent
  massless propagators; the dashed line corresponds to the photon propagator; the 2- and the 3-particle cut are
  indicated by the dashed red
  and blue lines, respectively; $k_1$, $k_2$ and $k_3$ are the loop momenta.}}\label{ACA4top}
\end{figure}

\begin{figure}[t]
\begin{center}
\vspace*{18mm}
\begin{tabular}{ccccc}
\begin{picture}(0,0)(0,0)
\SetScale{.8}
%
  \SetWidth{1.8}
\Line(-75,0)(-45,0)
\Line(45,0)(75,0)
  \SetWidth{.5}
\Line(-45,0)(45,0)
\DashCArc(-20,0)(25,-180,0){3}
%
%
\CArc(25,0)(20,-180,0)
\CArc(-20,0)(25,0,180)
%
%
  \SetWidth{1.8}
  \SetColor{Blue}
\DashLine(-45,-25)(5,25){3}
\end{picture}
&
\hspace{4cm}
&
\begin{picture}(0,0)(0,0)
\SetScale{.8}
%
  \SetWidth{1.8}
\Line(-75,0)(-45,0)
\Line(45,0)(75,0)
  \SetWidth{.5}
\Line(-45,0)(45,0)
\DashCArc(-20,0)(25,-180,0){3}
%
%
\CArc(25,0)(20,-180,0)
\CArc(0,0)(46,0,180)
%
%
  \SetWidth{1.8}
  \SetColor{Blue}
\DashLine(-45,-25)(25,45){3}
\end{picture}
&
\hspace{4cm}
&
\begin{picture}(0,0)(0,0)
\SetScale{.8}
%
  \SetWidth{1.8}
\Line(-75,0)(-45,0)
\Line(45,0)(75,0)
  \SetWidth{.5}
\Line(-45,0)(45,0)
\DashCArc(-20,0)(25,-180,0){3}
%
%
\CArc(25,0)(20,-180,0)
\CArc(-12,0)(34,0,180)
%
%
  \SetWidth{1.8}
  \SetColor{Blue}
\DashLine(-40,-25)(15,30){3}
\end{picture}
\end{tabular}
\end{center}
\vspace*{8mm}
\caption{{\sl Master Integrals for 3-particle cut of the topology displayed in Fig.~\ref{ACA4top}.}}\label{MIACA4}
\end{figure}
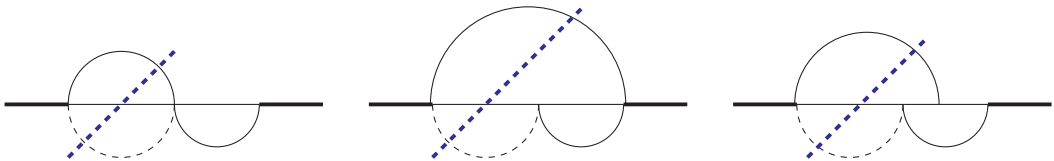

Let us consider the topology displayed in Fig.~{\ref{ACA4top}}, which corresponds to the last diagram shown in 
Fig~{\ref{abelianCA}}.
%
%
We carried out the reduction to
Master Integrals by means of  the package AIR \cite{Anastasiou:2004vj} and by means of a private code written by
one of us (in order to have a cross check).
In the diagram corresponding to the  topology in Fig.~{\ref{ACA4top}}, only a 2- and a 3-particle cut are
present (they are indicated by the dashed red and blue lines in Fig.~{\ref{ACA4top}}, respectively). 
We will first concentrate on the
3-particle cut; the 2-particle cut will be discussed at the end of this section.
Because of the delta function in Eq.~(\ref{energyfix}), one finds ($k_1=p_\gamma$)
\be
(p_b-k_1)^2 = m_b^2 (1-z) \, ;
\ee
therefore the corresponding propagator can be immediately factored out of the integrals. The reduction indicates
that the 3-particle cut of the topology in Fig.~{\ref{ACA4top}} has three Master Integrals, which are shown
in Fig.~{\ref{MIACA4}}.
%
%

Those Master Integrals have only single poles in $\epsilon$ (as long as $z\not=1$), and are sufficiently
simple to be calculated analytically by direct integration of a
suitable Feynman parameterization of the virtual loop and of the 3-particle phase-space.
By combining the output of the reduction to Master Integrals with the analytic
expressions of the latter, it is possible to find an expression for the
contribution of this particular cut to the photon energy spectrum.
It is then straightforward to check that the coefficient of the single pole in $\epsilon$, the finite
part, and the $O(\epsilon)$ term vanish in the $z \to 0$ limit, as expected from dimensional reasoning
\cite{Misiak:2008ss}. On the contrary, in the $z \to 1$ limit this particular cut shows divergences 
of the form $\ln^n(1-z)/(1-z)$ ($n=0,1$). Such divergences give origin to the plus distribution functions
which survive in the part of the photon energy spectrum which is proportional to $C_F^2$.
To see this one needs to extract a factor $(1-z)^{b\epsilon}$ (where $b$ is an arbitrary integer) out of each
Master Integral, to combine it with possible factors $1/(1-z)$ which emerge from the reduction procedure, and
then to replace the resulting expression by using
\begin{equation}
  (1-z)^{-1+b\epsilon} = \frac{1}{b\epsilon}\,\delta(1-z) +
  \sum_{n=0}^\infty\frac{(b\epsilon)^n}{n!}\left[\frac{\ln^n(1-z)}{(1-z)}\right]_+\,.
\end{equation}
The relation above explains why one needs to calculate the cofactor
of $(1-z)^{-1+b \epsilon}$ including terms of  $O(\epsilon)$:
the latter terms contribute to the  $O(\epsilon^0)$ part of the coefficient of $\delta(1-z)$.
For the definition of the plus distributions see Eqs.~(\ref{plusdis1}) and (\ref{plusdis2}).


\input{diagramsX}


Very often we had to deal with Master Integrals which we were not able to evaluate by
direct integration of their integral representation in terms of Feynman
parameters. A powerful tool to be used in these cases is the differential
equation method \cite{diffeq,Anastasiou:2002yz,Anastasiou:2003ds,Argeri}. The goal of
the method is to employ the output of the reduction procedure for a given
topology to build differential equations which are satisfied by the Master
Integrals of that topology. In our case, we consider differential equations with
respect to $z$, the only variable which appears in the Master Integrals. 
In order to generate these differential equations, it is necessary to take the
derivative of the integrand of a given Master Integral with respect to $z$. It
is interesting to observe that the only factor in the integrand which depends on
$z$ is the propagator associated to the delta function given in Eq.~(\ref{energyfix}).

A weakness of the differential equation method is the fact that there is no
general strategy which allows to fix the integration constants which are not
determined by the solution of the differential equations.   In a number of cases it was possible to fix the
missing constants by directly evaluating the Master Integral after setting $z=0$
from the start. However, we also found Master Integrals which are singular in
$z=0$; they appear in the reduction of the 3- and 4-particle cuts of the 3rd and
the last diagram given in 4th line in Fig.~\ref{abelianCA}. To fix the
integration constants in those cases we exploited the fact  that the coefficient
of each term in the $\epsilon$ expansion of each single cut contributing to the
photon energy spectrum must vanish in the $z \to 0$ limit. However, 
we had to deal with cuts for which this procedure did
not provide enough conditions to fix all of the unknown integration constants. Therefore we
calculated some of the Master Integrals for $z=1$ to reduce the number of
unknown integration constants; subsequently we fixed the remaining ones
by considering the $z \to 0$ limit of the cuts involving those integrals. Another procedure which we
employed to reduce the number of
unknown integration constants  consists in integrating some Master
Integral over $z$ from 0 to 1.  With these methods we were able to obtain analytic
expressions for all of the poles in $\epsilon$ appearing in the calculation of
the various cuts. 
For a few cuts we calculated the finite parts only numerically. Some of the
diagrams with non-trivial endpoint behavior were checked by an independent
calculation of the integrated spectrum. We would like to mention that for the
$(O_7,O_8)$-interference the endpoint singularities are only present in 3-particle
cuts of the $b$-quark self energies; 4-particle cuts are free of endpoint
singularities since the 4-particle phase space is proportional to $(1-z)$.

Now we will turn briefly to the evaluation of the 2-particle cut in
Fig.~{\ref{ACA4top}}. The reduction to Master Integrals is carried out  along
the lines of the reduction of the 3-particle cut discussed above. However, the
insertion of the delta function given in Eq.~(\ref{energyfix}) in the integrand 
is not necessary, since the
2-particle process takes place at fixed photon energy, $z=1$. We calculated  the
four Master Integrals which appear in this case by using a numerical method
based on sector decomposition \cite{Binoth:2000ps}; this technique  allows to
disentangle overlapping infrared, collinear and ultraviolet divergences. We
applied this numerical method in order to evaluate all of the Master Integrals
arising from 2-particle cuts. However, due to the presence of internal
thresholds, the integration over some of the Feynman parameters can only be
done numerically after a suitable contour deformation
\cite{Nagy:2006xy,Lazopoulos:2007ix,Anastasiou:2007qb}.

The infrared and collinear divergences appearing in the 2-particle cuts will
cancel after adding the 3- and 4-particle cuts, which also suffer from
infrared and collinear divergences. The remaining divergences are
ultraviolet. They are removed by adding counterterm diagrams with
appropriate $Z$-factor insertions (see App.~\ref{zfac}).


\section{\hspace{-3mm}\boldmath Estimating the numerical impact on Br($\bar B\to X_s\gamma$)}

In this section we investigate the numerical size of the
$(O_7,O_8)$-interference at $O(\alpha_s^2)$ at the level
of the branching ratio of the decay process
$\bar B\to X_s\gamma$. In order to do so, we adopt the notation
and conventions introduced in \cite{Misiak:2006ab}.
The $O(\alpha_s^2)$ correction to the function $K_{78}(E_0,\mu)$  in
Eqs.~(2.6) and (3.1) of \cite{Misiak:2006ab} is given by
\begin{align}\label{k782}
  K_{78}^{(2)}(E_0,\mu) &= \frac{C_F}{2}\int_{z_0}^1\!dz\,\bigg\{\widetilde Y^{(2)}(z,\mu) -
  C_F^2\frac{8\pi}{3}\alpha_\Upsilon\delta(1-z)\nonumber\\[2mm]
  &\hspace{3cm}-C_F\left[\frac{41}{2}-2\pi^2+12\,L_\mu\right]\widetilde Y^{(1)}(z,\mu)\bigg\}\,,
\end{align}
with $\alpha_\Upsilon=0.22$.
Note that Eq.~(\ref{k782}) refers to the $1S$-scheme for the
$b$-quark mass. The value used for this parameter in the numerics 
is $m_b^{1S}=4.68$\,GeV.

Combining the results of our present paper with those of \cite{Ewerth:2008nv} we are now able to
write down the
complete expression for $K_{78}^{(2)}$ containing all abelian and non-abelian contributions as well as
the effects of the masses of the $u,\,d,\,s,\,c$ and $b$ quarks running in the bubbles inserted in gluon lines.
This complete term affects the branching
ratio by an amount 
\begin{equation}\label{delBr}
  \Delta\text{Br}(\bar B\to X_s\gamma)_{E_\gamma>E_0} = \text{Br}(\bar B\to X_ce\bar\nu)_\text{exp}
  \left|\frac{V_{tb}V_{ts}^*}{V_{cb}}\right|^2\frac{6\,\alpha_{\rm em}}{\pi\,C}\,\Delta P(E_0)\,,
\end{equation}
where
\begin{equation}
  \Delta P(E_0) = 2C_7^{(0)\text{eff}}(\mu)C_8^{(0)\text{eff}}(\mu)\left(\frac{\alpha_s(\mu)}{4\pi}\right)^2
  K_{78}^{(2)}(E_0,\mu)\,,
\end{equation}
and $C$ is the so-called semileptonic phase-space factor. In order to compare with
Ref.~\cite{Misiak:2006ab}, we employ the numerical value for $C$ which was obtained from a fit of the
measured spectrum of the $\bar B\to X_cl\bar\nu$ decay in the 1S scheme\footnote{Using the
  numerical value for $C$ as obtained from a fit in the kinetic scheme raises the central value given in
  Eq.~(\ref{estimate}) by approximately 3\% \cite{Gambino:2008fj}.}
\cite{Bauer:2004ve,Hoang:2005zw}.

One might think that $\Delta\text{Br}(\bar B\to X_s\gamma)_{E_\gamma>E_0}$
in Eq.~(\ref{delBr}) simply represents the shift due to
$K_{78}^{(2)}$ of the theoretical prediction given in Eq.~(\ref{estimate}).
This is, however, not the case because an approximated
version of $K_{78}^{(2)}$ was already included in
\cite{Misiak:2006zs}: While the
$\beta_0$-part of $K_{78}$ (i.e. $K_{78}^{(2)\beta_0}$ when following the notation of
\cite{Misiak:2006ab}) was fully taken into account, the
remaining piece, $K_{78}^{(2)\rm rem}$, was calculated for large values
of $\rho=m_c^2/m_b^2$ and then interpolated (combined with
contributions not related to the $(O_7,O_8)$-interference) to the
physical value of $\rho$. To remove $K_{78}^{(2)\rm rem}$ from the
interpolation procedure and to replace it by the exact result obtained
by us, is beyond the scope of the present paper; this issue will be
correctly treated in a systematic update of Eq.~(\ref{estimate}) in the
near future.
To get nevertheless an idea of the numerical size of the $O (\alpha_s^2)$
contribution of the $(O_7,O_8)$-interference at the level of branching ratio for $\bar B
\to X_s \gamma$, we can ignore this issue and simply discuss a few numerical aspects of the quantity
$\Delta\text{Br}(\bar B\to X_s\gamma)_{E_\gamma>E_0}$, based on
$K_{78}^{(2)}$ which we calculated in this paper.

\begin{figure}[t]
  \begin{center}
    \epsfig{figure=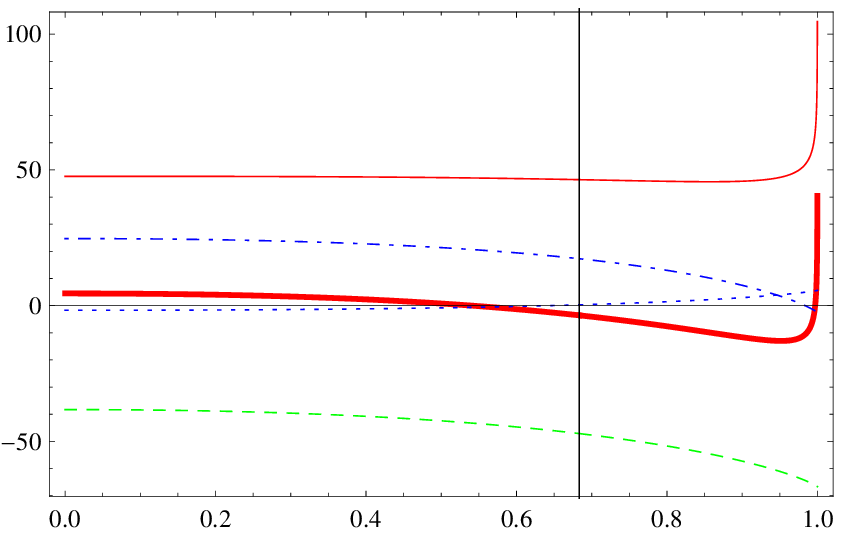,height=6cm}\\[-2mm]
    \hspace*{.6cm}{$z_0$}\\
    \hspace*{-10.2cm}\begin{rotate}{90}\hspace*{20mm}{
        $\Delta\text{Br}(\bar B\to X_s\gamma)\cdot 10^7$}\end{rotate}
    \caption{\sl $\Delta\text{Br}(\bar B\to X_s\gamma)_{E_\gamma>E_0}$ as a function of $z_0$. See text
      for more details.}\label{plot}
  \end{center}
\end{figure}

Fig.~\ref{plot} shows $\Delta\text{Br}(\bar B\to X_s\gamma)_{E_\gamma>E_0}$ as a function of $z_0$ for
$\mu=2.5\,\text{GeV}$, $\alpha_s(2.5\,\text{GeV})=0.271$, $C_7^{(0)\rm eff}(2.5\,\text{GeV})=-0.369$,
$C_8^{(0)\rm eff}(2.5\,\text{GeV})=-0.171$,
$N_L=3$ and $N_H=N_V=1$. (Note that the scale $\mu=2.5$\,GeV defines
the central value of the branching ratio given in Eq.~(\ref{estimate}).)
The remaining
numerical input parameters are taken from \cite{Misiak:2006ab}. The thick red solid line shows the
complete result, the thin red solid line  corresponds to the contribution proportional to $C_F^2$ (including the
numerical value of $C_F^2$), the green dashed line  
indicates the contribution proportional
to $C_FC_A$ (again including $C_FC_A$), and the blue dotted line is the contribution
stemming from massless and massive quark bubbles. 
The dash-dotted blue line indicates the corrections obtained by 
applying the 
large-$\beta_0$ approximation \cite{Brodsky:1982gc,Beneke:1994qe}.
The vertical line indicates the value of $z_0$ corresponding to the 
choice $E_0 = 1.6$ GeV.
It can be seen from the figure that the contributions proportional to $C_F^2$
and $C_FC_A$ cancel each other over almost the whole range of $z_0$, resulting in a contribution of size similar
to the one due to the fermionic corrections. Only in the region very close to the endpoint the
contribution proportional to $C_F^2$ dominates, due to its singular behavior for $z_0 \to 1$.
We stress, however, that the cancellations mentioned above refer to a value of
$\mu=2.5$ GeV; they do not occur anymore when going to
smaller values of $\mu$.

For a photon energy cut-off $E_0=1.6\,\text{GeV}$, as the one employed in Eq.~(\ref{estimate}), we find
the following numerical value for the quantity in Eq.~(\ref{delBr}) (using $\mu=2.5$ GeV):
\begin{align}
  \Delta\text{Br}(\bar B\to X_s\gamma)_{E_\gamma>1.6\,{\rm GeV}} &=
  \frac{C_F}{2}\,\bigl(52.21\,C_F-23.57\,C_A-2.66\,C_F^2-2.28\,N_L\nn\\
  &\qquad\qquad +8.73\,N_H -1.42\,N_V\bigr)\times 10^{-7} = -3.57\cdot 10^{-7}\, ,
\end{align}
where in the last step we inserted numerical values for the color factors and we
set $N_L=3$ and $N_H=N_V=1$. By comparing this with the central value of the
estimate given in Eq.~(\ref{estimate}), one sees that the $O(\alpha_s^2)$
corrections $K_{78}^{(2)}$ have an impact of
$-0.11\%$ at the level of the branching ratio. For the analogous effect generated by the large-$\beta_0$
approximation of $K_{78}^{(2)}$ only, $K_{78}^{(2)\beta_0}$, we find
$+0.56\%$. These results are, however, very strongly dependent on the
scale $\mu$.
For $\mu=1.25,\,2.34,\,5$\,GeV the corresponding numbers read 
$-5.15,\,-0.23,\,-0.07\%$ (full)
and $+0.10,\,+0.49,\,+1.15\%$ (large-$\beta_0$ approximation), respectively.
From these results we conclude that the large-$\beta_0$ approximation 
does not provide a good estimate of the 
full $O(\alpha_s^2)$ correction of the $(O_7,O_8)$-contribution for $\mu\in[1.25,5]$\,GeV.
As already mentioned, 
a more detailed analysis of the effect of the complete calculation of
$K_{78}^{(2)}$ on the central 
value of Eq.~(\ref{estimate}) would require 
to repeat the interpolation
procedure of \cite{Misiak:2006ab}. While this is beyond the scope of the
present work, we can conclude that the correction originating from the 
$(O_7,O_8)$-interference at $O(\alpha_s^2)$ will
not alter the central value of Eq.~(\ref{estimate}) by more than 1\%.


\section{\hspace{-3mm}Summary and conclusions}

In the present work we calculated the set of the $O(\alpha_s^2)$
corrections to the partonic decay process $b \to X_s \gamma$ which originates from the interference  of
diagrams involving the electromagnetic dipole operator $O_7$ with 
diagrams involving the chromomagnetic dipole operator $O_8$. These corrections are one of the elements
needed in order to
complete the calculation of the  branching ratio for the radiative
decay $\bar{B} \to X_s \gamma$ up to NNLO in QCD.

To carry out the calculation, we mapped the interference of diagrams
contributing to the processes $b \to s \gamma$, $b \to s \gamma g$, $b \to s
\gamma g g$ and $b \to s \gamma s \bar{s}$ onto 2-, 3- , and
4-particle cuts for the three-loop $b$-quark self-energy diagrams which
include insertions of the operators $O_7$ and $O_8$.
Subsequently, we evaluated each single cut by employing the Cutkosky rules.
From the technical point of view, the calculation was made possible by the 
use of the Laporta Algorithm \cite{Laporta:2001dd} to identify the needed Master Integrals, 
and of the differential equation method and sector decomposition
method to solve the Master Integrals. 

From the phenomenological point of view, it is  interesting to
estimate the effect of these corrections on the 
theoretical prediction
for the  
$\bar{B} \to X_s \gamma$ branching ratio. Our conclusion is that they
will not change its central value given in Eq.~(\ref{estimate}) by more than $1\%$.

At present, the largest theoretical uncertainty affecting the
prediction in Eq.~(\ref{estimate}) is of non-perturbative origin. It is expected to set a lower limit
of about $5\%$ on the total theoretical uncertainty for the prediction of the  $\bar{B} \to X_s \gamma$
branching ratio in the near future. The perturbative $O(\alpha_s^2)$ corrections of the $(O_7,O_8)$-interference
presented in this paper are a further contribution to make the perturbative uncertainty negligible with respect
to the non-perturbative one.


\section*{\normalsize Acknowledgments}
\vspace*{-2mm}

T.E.~would like to thank M.~Misiak for helpful discussions. A.F.~is grateful to J.~Vermaseren for his kind
assistance in the use of the 
algebraic manipulation program FORM \cite{Vermaseren:2000nd}, and to G.~Paz for  discussions.
The work of H.M.A.~was partially supported by ISTC A-1606 program.
T.E.~was supported by a European Community's Marie-Curie Research Training
Network under contract MRTN-CT-2006-035505 ``Tools and Precision Calculations for Physics Discoveries at Colliders''
(until October 2009) and by DFG through SFB/TR ``Computational Particle Physics'' (from October
2009 on).
C.G.~was partially supported by the Swiss National Foundation, by EC-Contract MRTN-CT-2006-035482
(FLAVIAnet) and by the Helmholz Association through funds provided to the virtual institute ``Spin and strong QCD''
(VH-VI-231); The Albert Einstein Center for Fundamental Physics is supported by the
``Innovations- und Kooperationsprojekt C-13 of the Schweizerische Universit\"atskonferenz SUK/CRUS''.
The work of G.O. was supported by the ToK Program ``Algotools'' MTKD-CD-2004-014319 (until June 2008) and in
part by NSF Grant No.~PHY-0855489.


\appendix

\section{\hspace{-3mm}Two-particle cuts} 

To obtain the contributions to the functions
$\widetilde Y^{(1)}$, $\widetilde Y^{(2,CF)}$ and
$\widetilde Y^{(2,CA)}$ (see Eqs.~(\ref{mastereq}) and (\ref{colorf})) originating from the 2-particle cuts
of the $b$-quark self energies discussed in the paper,
we first calculated the ultraviolet renormalized on-shell
matrix elements 
\[ 
\langle O_i \rangle \equiv
\langle  s \gamma |O_{i}|b \rangle\,, \qquad \mbox{($i=7,8$)} \, .
\]
For the
$(O_7,O_8)$-interference, $\langle O_7 \rangle$ and $\langle O_8 \rangle$
are needed to one-loop and two-loop accuracy, respectively. 
By considering only the  terms proportional to $C_F^2$ and $C_FC_A$ to $O(\alpha_s^2)$, which are the ones of
interest in order to obtain Eq.~(\ref{Yabelian}) and (\ref{Ynonab}), we write
\begin{equation}
\langle O_i \rangle= \langle O_{7} \rangle_{\rm{tree}}
\left[\delta_{i7}+ \frac{\alpha_s}{4\pi} \, C_F \, D_i^{(1)}
+ \left(\frac{\alpha_s}{4\pi}\right)^2 \, C_F \, \left(C_F D_{iF}^{(2)}+ C_A D_{iA}^{(2)} + \dots\right)
+ O(\alpha_s^3)\right].
\end{equation}
Note that the operator $O_7$ in $\langle O_{7} \rangle_{\rm{tree}}$
contains the $b$-quark running mass  $\overline m_b(\mu)$.
For $O_7$ the relevant results are  \cite{Blokland:2005uk,Asatrian:2006ph}
\begin{eqnarray}
  D_7^{(1)}&=&-\frac{1}{\epsilon^2}-\frac{1}{\epsilon}\left(2\,L_\mu+2.5\right) -2 \,L_\mu^2-7\,L_\mu -
  6.8225\nonumber\\[1mm]
  &&-\,\epsilon\,\left(1.3333 \,L_\mu^3 + 7 \,L_\mu^2 + 13.6449 \,L_\mu +13.4779\right)\nonumber\\[2mm]
  &&-\,\epsilon^2\,\left(0.6667 \,L_\mu^4+ 4.6667 \,L_\mu^3 + 13.6449 \,L_\mu^2 + 26.9559 \,L_\mu + 26.1412\right), 
\end{eqnarray}
and for  $O_8$ one finds
\begin{eqnarray}
  D_{8}^{(1)}&=&2.6667 \,L_\mu + 1.4734 + 2.0944\,i\nonumber\\[2mm]
  &&\quad+\,\epsilon\,\left[2.6667 \,L_\mu^2+ 2.9468 \,L_\mu-1.1947+\,i\,(4.1888\,L_\mu + 4.1888)\right]
  \nonumber\\[2mm]
  &&\quad+\,\epsilon^2\big[1.7778\,L_\mu^3+2.9468 \,L_\mu^2-2.3894 \,L_\mu -5.5373\nonumber\\[2mm]
    &&\hspace{1.8cm}+\,i\,\left(4.1888 \,L_\mu^2+8.3776 \,L_\mu +2.1627\right)\big],\nonumber\\[3mm]
  D_{8F}^{(2)}&=&
  D_8^{(1)}\left(-\frac{1}{\epsilon^2}-\frac{1}{\epsilon}\left(2\,L_\mu+2.5\right)\right)-5.3333 \,L_\mu^3
  -32.2802 \,L_\mu^2-50.9612 \,L_\mu\nonumber\\[1mm]
  &&\quad -1.8875-i\left(4.1888 \,L_\mu^2+31.4159 \,L_\mu +29.8299\right),\nonumber\\[3mm]
  D_{8A}^{(2)}&=&15.111 \,L_\mu^2+31.6617 \,L_\mu + 2.38332 + i\left(23.7365 \,L_\mu +28.0745\right)\,.
\end{eqnarray}
Taking into account the phase-space factors in $d = 4 -2 \epsilon$ dimensions, we easily obtain the contributions
to the functions $\widetilde Y^{(1)}$, $\widetilde Y^{(2,\mbox{{\tiny CF}})}$ and
$\widetilde Y^{(2,\mbox{{\tiny CA}})}$ which originate from  the 2-particle cuts only:
\begin{align}
  \widetilde Y^{(1)}_{\rm 2-cuts}(z,\mu) &= \left[ \frac{2}{9} \left(33 - 2\pi^2\right)
    + \frac{16}{3} \,L_\mu \right]\,\delta(1-z)\,,\nonumber \\[2mm]
  \widetilde Y^{(2,\mbox{{\tiny CF}})}_{\rm 2-cuts}(z,\mu) &= \bigg[
    \frac{1}{\epsilon^2}\left(-5.89368 - 10.6667 \,L_\mu\right) +
    \frac{1}{\epsilon}\left(-15.849 - 72.6954 \,L_\mu - 53.3333 \,L_\mu^2\right)\nonumber \\
    &\hspace{1cm} + 3.01591- 246.244 \,L_\mu - 335.42 \,L_\mu^2 - 135.111 \,L_\mu^3\bigg]\,\delta(1-z)\,,
  \nonumber \\[2mm]
  \widetilde Y^{(2,\mbox{{\tiny CA}})}_{\rm 2-cuts}(z,\mu) &= \left[
    4.76664 + 63.3235 \,L_\mu + 30.2222 \,L_\mu^2 \right]\,\delta(1-z)\,.
\end{align}

\noindent The results given in this appendix were already used by one of us in \cite{Ali:2007sj}.


\section{\hspace{-3mm}Renormalization constants}\label{zfac}

In this appendix, we collect the  explicit expressions of the renormalization constants needed for the
ultraviolet renormalization in our calculation.
\noindent The strong coupling constant is renormalized in the $\MS$ scheme:
\be\label{alpha_ms}
 \alpha_s^{\rm bare} =
 \mu^{2\epsilon}\left(\frac{e^\gamma}{4\pi}\right)^\epsilon
 Z_\alpha^{\MS}\,\alpha_s(\mu)\,,
\ee
with
\be
 Z_\alpha^{\MS} = 1 - \frac{1}{\epsilon}\left(\frac{11}{3}\,C_A-\frac{4}{3}\,T_RN_F\right)
 \frac{\alpha_s(\mu)}{4\pi} + O(\alpha_s^2)\,,
\ee
and $N_F=N_L+N_H+N_V$. The $b$-quark mass which appears  in the operators $O_{7,8}$, as well as  the Wilson
coefficients $C_{7,8}^{\rm eff}$ themselves, are also renormalized in the $\MS$ scheme  \cite{Misiak:1994zw}:
\begin{align}
 Z_{m_b}^{\MS} &= 1 - \frac{3\,C_F}{\epsilon}\frac{\alpha_s(\mu)}{4\pi} + O(\alpha_s^2)\,,\nonumber\\[2mm]
 Z_{77}^{\MS} &= 1 + \frac{4\,C_F}{\epsilon}\frac{\alpha_s(\mu)}{4\pi} + O(\alpha_s^2)\,,\nonumber\\[2mm]
 Z_{87}^{\MS} &= - \frac{4\,C_F}{3\,\epsilon}\frac{\alpha_s(\mu)}{4\pi} +
 \bigg\{\frac{1}{\epsilon^2}\left(\frac{34}{9}\,C_FC_A-8\,C_F^2-\frac{8}{9}\,C_FT_RN_F\right)\nonumber\\[2mm]
 &\hspace{1cm} + \frac{1}{\epsilon}\left(-\frac{101}{27}\,C_AC_F+\frac{8}{3}\,C_F^2 +
 \frac{28}{27}\,C_FT_RN_F\right)\bigg\}\left(\frac{\alpha_s(\mu)}{4\pi}\right)^2 + O(\alpha_s^3)\,,\nonumber\\[2mm]
 Z_{88}^{\MS} &= 1 + \frac{2}{\epsilon}\left(4\,C_F-C_A\right)\frac{\alpha_s(\mu)}{4\pi} + O(\alpha_s^2)\,.
\end{align}

\noindent All the remaining fields and parameters are 
renormalized in the on-shell scheme. The on-shell renormalization constant for
the $b$-quark mass is given by
\be
 Z_{m_b}^{\rm OS} =
 1-C_F\,\Gamma(\epsilon)\,e^{\gamma\epsilon}\,
 \frac{3-2\epsilon}{1-2\epsilon}
 \left(\frac{\mu}{m_b}\right)^{2\epsilon}\frac{\alpha_s(\mu)}{4\pi} +
 O(\alpha_s^2)\, .
\ee
The renormalization constants for the gluon field and the $s$- and 
$b$-quark fields are
\begin{align}
  Z_3^{\rm OS} &=
   1-\frac{4}{3}\,T_R\left(N_H+N_V\rho^{-\epsilon}\right)\Gamma(\epsilon)\,e^{\gamma\epsilon}
   \left(\frac{\mu}{m_b}\right)^{2\epsilon}
   \frac{\alpha_s(\mu)}{4\pi} + O(\alpha_s^2)\,,\nonumber\\[2mm]
  Z_{2s}^{\rm OS} &= 1 + O(\alpha_s^2)\,,\nonumber\\[2mm]
  Z_{2b}^{\rm OS} &= 1 - C_F\,\Gamma(\epsilon)\,e^{\gamma\epsilon}\,\frac{3-2\epsilon}{1-2\epsilon}
  \left(\frac{\mu}{m_b}\right)^{2\epsilon}\frac{\alpha_s(\mu)}{4\pi} + O(\alpha_s^2)\,,
\end{align}
where $\rho=m_c^2/m_b^2$.


\end{document}